\documentstyle[prl,psfig,aps]{revtex}
\def\vf#1#2{\hbox{$V_{#1#2}$} }
\def\zm#1#2{\hbox{$Z_{#1#2}^{-1}$} }
\def\kf#1{\hbox{$k_{F,#1}$} }

\def\U#1{{u}_{#1}} 

\def\dvf#1{\hbox{$\Delta V_{#1}$} }
\def\sof{\hbox{${SO(5)}$} }
\def\soft{\hbox{${\other{SO}(5)}$} }

\def\other#1{\widetilde{#1}}

\def\rsof{\hbox{${\cal R}$} }
\def\rsoft{\hbox{${\other{\cal R}}$} }

\def\dz#1{\hbox{$\Delta Z^{-1}_{#1}$} }

\newcommand{\OO}{{\cal O}}

\newcommand{\rr}{\nu}

\def\flow{\tau}

\def\tp{\hbox{$t_{\perp}$}}

\def\v#1{{\bf #1}}

\def\iom{i \omega}
\def\om{\omega}
\def\kx{{k_{\parallel}}}
\def\ky{{k_{\perp}}}

\def\kom{\v k}
\def\koms{\v k \g}

\def\g{\sigma}

\def\typ#1{^{(#1)}}
\def\tgen{\typ{}}

\def\wtilde{\widetilde}
\def\wbar{\overline}

\def\Lam{\Lambda}

\long\def\taglia#1{}

\long\def\tagliasi#1{}

\def\up{\uparrow}
\def\down{\downarrow}

\def\beq{\begin{equation}}
\def\eeq{\end{equation}}
\def\beqn{\begin{eqnarray}}
\def\eeqn{\end{eqnarray}}

\def\eqref#1{ %
 (\ref{#1})}

\long\def\singlecol#1{
\twocolumn[\hsize\textwidth\columnwidth\hsize\csname @twocolumnfalse\endcsname
              #1]}
\long\def\beginfigeps#1#2{
\begin{figure}[htb] 
\vspace*{-.4cm}
   \centerline{\psfig{file=#1,width=8.4cm}}
    \protect{#2}
\vspace*{-.3cm}
 \end{figure}
}
\long\def\taglia#1{}

\def\figa{\beginfigeps{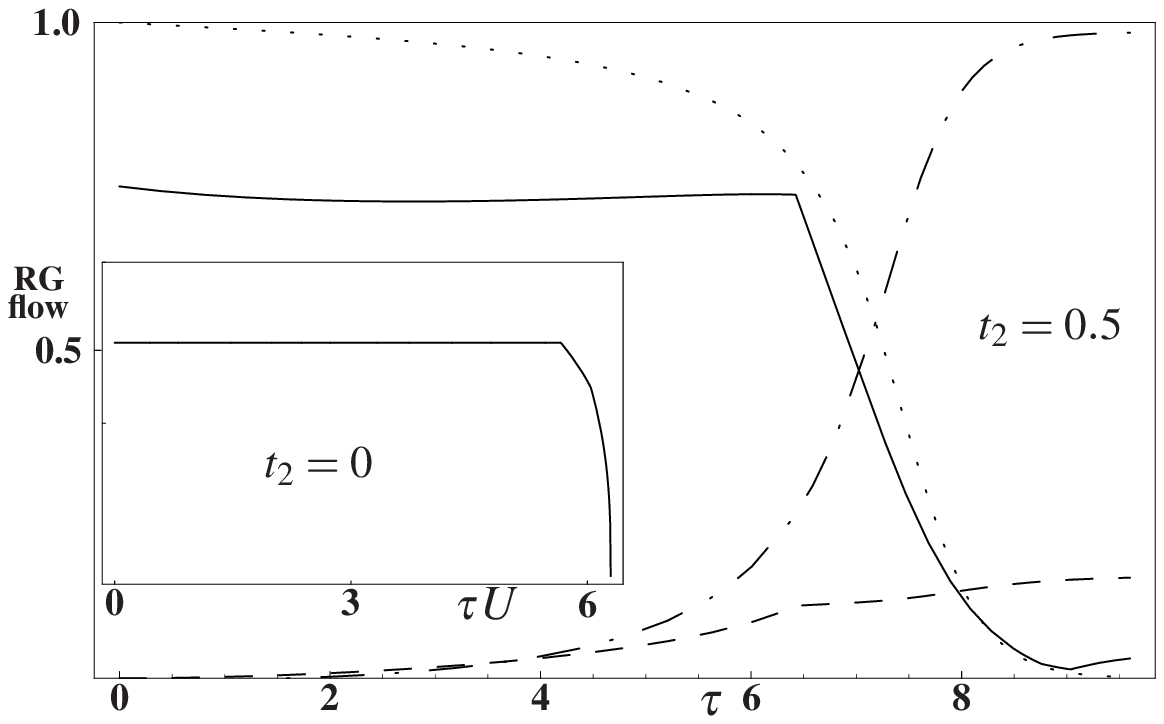}{\vspace*{-.4cm}
\caption{\label{figa}
RG flow of the \sof-breaking terms
$3g_1\typ{max}/g_0\typ{max}$ (full),  
$3 g_{ph}\typ{max}/g_0\typ{max}$ (dashed), 
$\dvf{\flow}/\dvf{\flow=0}$ (dotted), and
$\dz{\flow}$ (dash-dotted),
 as function of $\flow=-\log(\Lam/\Lam_0)$ for  $U=1$,
$t=\tp=1$, $t_2=0.5$ and half filling.
Here, $\dvf{\flow}=\vf{0}{\flow}-\vf{\pi}{\flow}$, and
$\dz{\flow}\equiv\left((\zm{0}{\flow}/\zm{\pi}{\flow})^2-1\right)/
 \left(\vf{0}{\flow=0}/\vf{\pi}{\flow=0}-1\right)$.
The inset shows  $3 g_1\typ{max}/g_0\typ{max}$ vs $\flow U$ for $t_2=0$.
}\vspace*{-.0cm}
}}
\def\figb{\beginfigeps{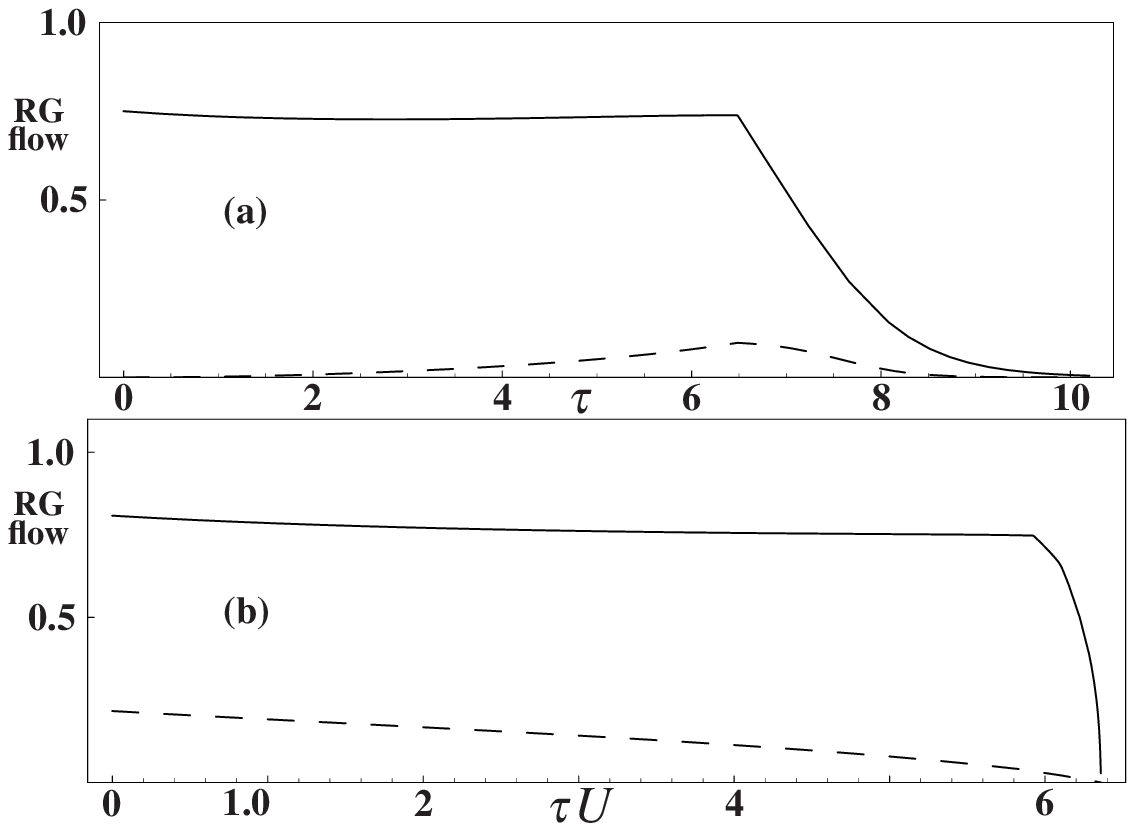}{\caption{\label{figb}
\soft-breaking couplings 
$3 \wtilde g_1\typ{max}/\wtilde g_0\typ{max}$ (full) and
$3 \wtilde g_{ph}\typ{max}/\wtilde g_0\typ{max}$ (dashed)
as a function of $\flow$.
(a) shows  the results of the RG procedure (i) and
(b) shows the results of (ii)  as defined in the text.
The parameters are as in Fig. 1.
}}}

\begin{document}

\draft   %

\title{ 
Renormalized \sof symmetry in ladders with
 next-nearest-neighbor hopping
}

\author{E. Arrigoni and W. Hanke
}
\address{ 
Institut f\"ur Theoretische Physik,
Universit\"at W\"urzburg,
D-97074 W\"urzburg, Germany
}

\singlecol{

\date{{\bf to appear in Phys. Rev. Lett.} Mar. 99}

\maketitle

\begin{abstract}

We study the  occurrence of \sof symmetry 
in the low-energy sector of 
  two-chain Hubbard-like systems  by analyzing the flow of the running 
couplings ($g$-ology) under
  renormalization group in the weak-interaction limit.
It is shown that \sof  is  asymptotically restored for low energies 
for rather general parameters of the bare Hamiltonian.
This holds also with inclusion of a
 next-nearest-neighbor hopping which explicitly breaks particle-hole symmetry
provided one
 accounts for a different single-particle weight for
 the quasiparticles of the two bands of the system.
The physical significance of this renormalized \sof symmetry is discussed.

\end{abstract}

\pacs{PACS numbers : 
71.10.-w  %
11.10.Hi  %
11.30.Ly  %
}


} %

Recently, it was shown that both the two-dimensional (2-D) Hubbard and
the $t-J$ model enjoy an approximate \sof symmetry
\cite{me.ha.97,ed.ha.97},
unifying antiferromagnetism (AF) with d-wave superconductivity
(dSC)\cite{zhan.97}.
This symmetry principle gives a definite microscopic description of
the AF $\to$ dSC transition as the chemical potential is varied. From
the \sof multiplet structure verified by exact cluster
diagonalization, one can see how the \sof superspin vector is rotated
from AF to dSC direction and show that at the critical chemical
potential, the energy barrier $\Delta E$ between AF and dSC states 
is an order of
magnitude smaller ($\sim J/10$) than the exchange coupling $J$, i. e.,
the natural parameter in the model. This finding is clearly of
importance: while it is well-established that both $t-J$ and Hubbard
models reproduce very successfully the ``high'' and ``medium''-energy physics 
of order $U$ and $J\sim t^2/U$ of the cuprates, the low-energy
content of order of the superconducting gap  has so far eluded 
theoretical investigations. The variance $\Delta E$ of the multiplet 
splitting is a
well-defined measure of ``how-good'' the \sof symmetry is realized in
the bare model considered for a given system size. It seems then a
natural question to ask, whether this deviation from exact \sof
symmetry  remains small or even vanishes if one goes to the
infinite-volume limit and to lower
energies, i. e. under renormalization-group (RG) flow.

{
In this Letter, 
 we demonstrate 
that the low-energy regime of
 rather general Hubbard-type
models including finite-ranged interactions (provided they are weak)
between one and two dimensions, i. e. ladders\cite{ladders}, 
is indeed dominated by 
the scaling 
towards an  \sof-invariant  model. 
This is a remarkable result, since it is the first model,
which is non-\sof invariant at the bare starting (microscopic) level,
where the existence of  \sof symmetry is proven for 
low energies\cite{scalingmeans}.
We first 
show that this holds  for the {\it particle-hole}- (ph) symmetric case with
nearest-neighbor hopping only\cite{fisher}.
}
In addition, we
consider the effect of 
a next-nearest-neighbor (intrachain) hopping 
$t_2$  which 
produces
an explicit breaking of the ph
 symmetry. 
We demonstrate that also in this case the system becomes \sof symmetric
for low energies
(at least up to order $t_2^2$)
provided one considers a 
 {\it renormalized} 
 \sof transformation\cite{soft}, 
which takes into account a different renormalization of the
single-particle weight of the bonding and antibonding bands.
This  result is of  importance because of two points: (i)
it sheds light on the
effect of the longer-range hoppings 
in general 
  on the fate of \sof symmetry. This issue is
also under intensive discussion 
in the case of 2-D systems
\cite{ed.ha.97,henl.97,zhan.rece.97}: {
$t_2$ is known to strongly affect AF correlations and the
Fermi-surface topology in the cuprates.}
(ii) In addition,
the renormalized \sof representation introduced here for the first time,
is likely to be realized in a significantly larger class of physical
systems 
{
allowing, 
for example, for asymmetries in the AF- and dSC- phases, such as
different transition temperatures.}

Specifically, we consider  two coupled chains in the band
representation with total 
``low-energy''
action $S=S_0 + S_I$, where the non-interacting
 part $S_0$ 
can be written at a given point $\flow$ in the RG flow as
\beq
\label{s0L}
S_0 = 
\sum_{\kom,\g}
C^{\dag}_{\koms} \ \zm{\ky}{\flow}
\left[ \iom - \rr \ \vf{\ky}{\flow}\  \kx \right] 
\ C_{\koms}  \;.
\eeq
Here,
 $C_{\koms}$  ($C^{\dag}_{\koms}$) are Grassmann variables
associated with the destruction (creation) of a fermion, and $\g$ is the
spin. 
$\kom\equiv \{\iom,\ky,\kx,\rr\}$ 
is a shorthand notation for the 
Matsubara frequency $\iom$, and
the momentum perpendicular ($\ky=(0,\pi)$) and
 parallel
 ($\kx$) to the chain direction. The latter momentum is measured relative
 to  the
 Fermi point $\rr\ \kf{\ky}$
 associated with the ``band'' $\ky$ with Fermi velocity
 $\vf{\ky}{\flow}$, 
and 
$\rr=\pm1$ refers to right- and left-moving fermions, respectively.
This action is 
restricted to modes with $|\kx|<\Lam$ with $\Lam=\Lam_0 e^{-\flow}$.
{
This weak-coupling RG method has previously been applied
to obtain the
phase diagram of the two-chain Hubbard model\cite{ladders}.
}
In order to study the occurrence of \sof symmetry in Hubbard-like
models, 
it is
convenient to rewrite the interaction part $S_I$ of the action  in terms of
\sof-invariant and \sof-breaking terms { \protect{\cite{shelton}} }.
Defining the \sof spinor as in Ref. \onlinecite{notation} 
\beq
\label{spinor}
 \Psi_{\kom} \equiv 
\left\{
 C_{\kom\up},
C_{\kom\down},
-\cos\ky  C^{\dag}_{\wbar \kom\up },
 -\cos \ky C^{\dag}_{\wbar \kom\down}
\right\}^t
\;,
\eeq
where
$\wbar\kom$ stands for $ \{-\iom,-\ky+\pi,-\kx,\rr\}$), 
the interacting action $S_I$ can be 
shown to be expressable in terms of
the $4 \times 4$ charge-rotation Dirac matrix $\Gamma^{15}$ \cite{notation}:
\beqn
\nonumber
S_I  && =
 \frac{1}{2 \beta  N}\sum_{\kom_1, \cdots \kom_4}{}^{'} g_0(\cdots)  \ 
\Psi^{\dag}_{\kom_1} \left(1 + a(\cdots) \Gamma^{15} \right)  \Psi_{\kom_2} 
\\
\label{si}
\times &&
\Psi^{\dag}_{\kom_3}  \left(1 + b(\cdots) \Gamma^{15} \right)
 \Psi_{\kom_4} 
+  (k_1,k_3) \leftrightarrow (k_2,k_4)
\;.
\eeqn
Here,
 $(\cdots)$ represent the sets of variables on which the
 couplings $g_0$, $a$, and $b$ depend.
As usual,
 each coupling  can be considered as 
dependent only on the Fermi momenta closest to where the corresponding
 process takes
 place, i. e. the $(\cdots)$ are labeled by
$(\rr_1,\ky_1;\rr_2,\ky_2|\rr_3,\ky_3;\rr_4,\ky_4)$.
Moreover,  $\sum'$ denotes a sum with conservation of 
  frequency and lattice momentum.

The \sof-symmetric part $S_I^{(0)}$ of $S_I$ is given by Eq. \eqref{si}
 with $a(\cdots)=b(\cdots)=0$\cite{scalar,sc.zh.97}.
For a general \sof-invariant action 
$S_I^{(0)}$ it can be shown that
one can restrict oneself to 
the  seven independent couplings 
 $g_0\typ4$, $g_0\typ{4;0-\pi}$, $g_0\typ{2}$, $g_0\typ{2;0-\pi}$,
$g_0\typ{1}$, $g_0\typ{1t,0-\pi}$, and $g_0\typ{1,0-\pi}$,
defined in 
analogy with the $g$-ology formalism \cite{coupl,soly.79,web}.
The \sof-breaking term $S_I^{(1)}$ 
of the
 interacting action 
with ph symmetry
can be demonstrated to be
the term proportional to
 $g_0(\cdots) a(\cdots) b(\cdots)$ in \eqref{si} 
and thus we define the corresponding
 \sof-breaking couplings as $g_1(\cdots) \equiv g_0(\cdots)\ a(\cdots)\ b(\cdots)$.
In this case, we can restrict ourselves to only $5$ 
 independent couplings (it can be shown that the others are redundant), namely,
$g_1\typ4$, 
$g_1\typ2$, $g_1\typ1$, $g_1\typ{1t;0-\pi}$, and $g_1\typ{1;0-\pi}$\cite{coupl}. 
At half-filling and with $t_2=0$, 
 the Hamiltonian is ph symmetric, since
 the velocities of the two bands $\vf{\ky=0}{}$ and $\vf{\ky=\pi}{}$
 are equal.
Since the ph-breaking terms in \eqref{si} are proportional to
$\Gamma_{15}$, one can set in the ph-symmetric case
 $a(\cdots)=-b(\cdots)$, and 
consider the RG flow of the
couplings $g_0\tgen$ and $g_1\tgen$ only.

To begin with, we have evaluated
the RG
equations 
for the $g_0\tgen$ and 
$g_1\tgen$
 couplings \cite{web}
at one loop  by  using the 
standard $g$-ology procedure 
 (cf. Refs. \onlinecite{soly.79,ladders}),  including the
interband umklapp processes.
As already shown for the two-chain case \cite{ladders},
the system always flows to strong coupling, i.e. the $g$'s
{
 diverge at a value of $\flow=\flow_c \propto 1/g(\flow=0)$, 
$g(\flow)$ being the scale of the  interaction 
(proportional to the maximum of all $g_i\tgen(\flow)$).}
This
 signals an
 instability towards some gapped state.
Nevertheless, the striking new result  is that 
even in a non-\sof-invariant system, like, e.g., the Hubbard model,
 the \sof-invariant couplings $g_0\tgen$ dominate with
respect to the symmetry-breaking couplings $g_1\tgen$, 
when approaching $\flow_c$ \cite{controlled}. 
This can 
be seen from the ratio of the maxima 
of these two types of couplings, $g_1\typ{max}(\flow)/g_0\typ{max}(\flow)$
going to zero, as shown 
in  the inset
 of Fig. \ref{figa}.
Here, 
$g_i\typ{max}(\flow)$ is defined as the largest absolute value, and
thus the scale
of the couplings of a given type $i$ ($i=0,1,ph$) at a given $\flow$.
This result 
 implies that {\it the low-energy modes of the system can be
described by an effective \sof-symmetric action, at least for sufficiently small
$g(\flow=0)$} 
\cite{controlled,scalingmeans}.
In fact, we have verified that
this  occurs 
for very general values of the 
Hamiltonian, including longer-ranged interactions.

A next-nearest-neighbor hopping $t_2$ breaks ph symmetry
explicitly and requires the introduction of a ph breaking
interaction $S_I^{(ph)}$. 
Here,
$a(\cdots) \not= -b(\cdots)$  in Eq. \eqref{si}, 
and thus we need extra couplings 
$g_{ph}(\cdots)$, which we have defined 
as $g_{ph}(\cdots)=g_0(\cdots) \ \left( a(\cdots) + b(\cdots) \right)/2$.
In this case, one can show that 
the  couplings can be restricted to
 $g_{ph}\typ4$, $g_{ph}\typ2$, and $g_{ph}\typ1$\cite{coupl,web}.
 The initial ($\flow=0$) source of ph-symmetry
 breaking for $t_2\not=0$ stems from the
non-interacting part of the action $S_0$, due to the difference of the Fermi
velocities $\dvf{0}$ of the two bands. 
In the following, we will show that 
\sof symmetry is restored 
(at least up to $\OO(t_2)^2$),
at low energies,  even in the presence of this ph-
(and thus \sof-)breaking term. 

\vspace*{-.5cm}
\figa

These results are obtained 
on the basis of two complementary RG calculations. Calculation
(i)  considers the RG flow of the
self-energy parameters 
$\vf{\ky}{\flow}$ and $\zm{\ky}{\flow}$ at two loops, and of the
coupling parameters $g_i\tgen(\flow)$ at one loop, 
taking
the  
$\flow$ dependence of all the
 parameters at each RG step fully into account. \label{selfcons}
This first calculation (i), although not rigorously controlled (see below),
 is motivated by the fact that
we are  interested in 
 studying the RG flow of the
self-energy, which is the leading symmetry-breaking term when $t_2$ is
included 
\cite{web}.
In a second calculation (ii),
 we will show how our main results about  the renormalized \soft
symmetry obtained within this first procedure
can be achieved also in an
alternative, more controlled
way, where we consider only the renormalization of the $g_i\tgen$.
Nevertheless, the first calculation (i) 
is instructive,
in order to provide a
 physical interpretation for the single-particle renormalization factors
 $Z$ as discussed in the conclusions.
Indeed,  in  procedure (i)  the $Z$ factors, and thus the
renormalized \sof transformation, derive naturally
from the RG
flow, while in (ii) they are introduced right at the outset.
\label{motivation}

In calculation (i),
the relevant
 part of the renormalized action has the
form of Eq.~\eqref{s0L} with
$\flow$-dependent 
Fermi velocities and 
 single-particle weights.
{\noindent The flow of these parameters is shown in 
Fig. \ref{figa}.}
As  the bare ($\flow=0$) Hamiltonian,  we take the half-filled Hubbard ladder 
with isotropic
 intrachain and interchain hoppings 
$t=\tp=1$, 
 next-nearest-neighbor hopping $t_2=0.5$ 
(corresponding to $\dvf{\flow=0}\approx 1.9$), and $U=1$\cite{U}.
Actually, we have verified that 
the results we are discussing below are  rather general and hold
also in the presence of anisotropy $\tp \not=t$ and  nearest-neighbor 
interactions ({$\gtrsim -U$}).
The $t_2=0$ case\cite{fisher}, discussed above, is plotted in the inset
for comparison.
 
 For the $t_2\not=0$ case, 
$\dvf{\flow}$ (dotted line) flows to zero, 
 but $\dz{\flow}$
(dash-dotted line, initially zero) scales to unity.
Therefore, the initial asymmetry between the bands due to the
different Fermi velocities is transferred into a difference in the
single-particle weights $Z$, such that for large $\flow$
$\zm{0}{\flow}/\zm{\pi}{\flow} \to
\sqrt{\vf{0}{\flow=0}/\vf{\pi}{\flow=0}}$. 
\label{standard}
In order to restore the coefficient of the $\iom$ term in \eqref{s0L} 
to unity, the standard procedure\cite{soly.79} is to
reabsorb this renormalization into the definition of new
Grassmann variables $\wtilde C_{\koms}$ and to set 
$  \sqrt{\zm{\ky}{\flow}} C_{\koms}=  \wtilde C_{\koms}$. 
This standard procedure
is dictated by the requirement
to  identify the 
{\it canonical} Fermi operators with correct anticommutation
relations, as 
 will be discussed at the end.
\figb
In this way,
 the non-interacting 
part of the (renormalized)  action will again be  symmetric under
exchange of the two bands (and thus \sof symmetric in the new fields).
This transformation, however, also affects the interaction part, and one
should consistently 
 redefine the renormalized
\sof spinor  
 in  \eqref{spinor} to $\wtilde \Psi_{\kom}$, whereby  the $C_{\koms}$
 are again replaced
with 
the $\wtilde C_{\koms}$.
The couplings defined in this way are of course different from the
original ones and we will distinguish them with a tilde, i.e.
$g_i\tgen \to \wtilde g_i\tgen$. 
The remarkable result is 
that {\it the transformation
which makes the  non-interacting part of the action
\sof-symmetric also restores \sof in the  interacting part}.
This is demonstrated
in  Fig. \ref{figb}a, which plots
 the ratio of the $\wtilde g_i\typ{max}$
as a function of the flow parameter $\flow$. 
We note that
the non-\sof couplings $\wtilde g_1\typ{max}$ and $\wtilde
g_{ph}\typ{max}$ all flow to zero (relative to the $\wtilde
g_0\typ{max}$). 
Thus,
at
large $\flow$,
\sof symmetry 
is  restored for low energies
 in the
``~$\wtilde{\ }$~'' basis.
However, at the energy scale
where $\dvf{\flow}$ starts to decrease and
$\dz{\flow}$ starts
to become finite ($\flow\sim 7$ in Fig. \ref{figa}), 
the renormalized couplings can be shown 
to become large
 and the
weak-coupling expansion is no longer controlled, as anticipated.

To support this
physically appealing yet 
uncontrolled calculation, 
we  verify,
in terms of a {\it controlled} RG calculation \cite{controlled}
(i. e., at one loop),
that \soft  symmetry is indeed recovered
at least up to $\OO(t_2^2)$.
This alternative derivation 
clarifies
why the single-particle weights renormalize  proportionally to
$(\sqrt{\vf{}{\flow=0}})^{-1}$, as 
obtained asymptotically in the two-loop calculation.
\label{hatT}
In this RG procedure (ii),
we start from the action $S_0+S_I$ and carry out the
transformation $\other T$ on the Grassmann variables right {\it at the
  outset},
 where $\other T$ is defined as
$ \other T C_{\iom,\kx,\ky}= 1/\sqrt{\U{\ky}} \other C_{\iom',\kx',\ky }$
with 
$\iom'=\iom/\U{\ky}$ and $\kx'=\kx \ \U{\ky}$, and
$\U{\ky}=\sqrt{\vf{\ky}{}}$ \cite{scaling}.
Such a transformation, which  is always possible with Grassmann variables,
  is motivated by our first calculation (i).
By changing  the sum over $\iom$ and $\kx$ into a sum over 
$\iom'$ and $\kx'$ separately for each band, the
non-interacting part of the action again recovers its explicit \sof
symmetry.
Furthermore,
by defining a new \sof spinor $\other \Psi_{\kom}$ in terms
 of the $\other C$, we obtain  new couplings 
$\other g_0\tgen$, $\other g_1\tgen$, and $\other g_{ph}\tgen$, as in
  step (i). 
In Fig. \ref{figb}b, we show the corresponding RG flow of the ratios of the 
$\other g_i\typ{max}$.
With 
increasing RG parameter $\flow$ the ph-symmetry breaking term 
$\other g_{ph}\typ{max}/\other g_0\typ{max}$ vanishes (full line), while the \soft-breaking term
$\other g_1\typ{max}/\other g_0\typ{max}$ 
goes to 
a finite but rather small value \cite{approx} (dashed line).
The \soft symmetry  thus is recovered up to a very high degree of
precision for low energies\cite{scalingmeans}.
In contrast with the results of 
procedure (i), the result of (ii)
is {\it controlled} for small $g(\flow=0)$ \cite{controlled,remaining}.
In this way, we have  shown {\it in a controlled way} 
that \soft is restored 
for low energies
{\it 
at least} up to order $t_2^2$ for small $g(\flow=0)$.
Our  two-loop calculation (i)
further suggests that even
this \soft-breaking term
of order $t_2^2$ 
 might be
 removed by the self-energy renormalization.

The renormalized 
\soft symmetry introduced here, and the related  
  renormalization 
of the single-particle weights $Z_{\ky\flow}$,
can  be understood 
in terms of a simplified scheme, which  renormalizes the
Hamiltonian,
by  restricting the Hilbert space
to a subspace with energy
 $\om$ smaller
than a certain cutoff $\om_0\propto\Lam_0\exp(-\flow)$. 
In the restricted subspace,
the total integrated spectral density
(which we identify with  $Z_{\ky\flow}$) 
will be less than one.
Since 
the spectral sum rule 
identifies
$Z_{\ky\flow}$
with the anticommutator 
of the Fermi operators $C_{\kom}$,
the canonical Fermi operators with anticommutator equal to $1$ in
this subspace are the transformed field operators 
$\wtilde C_{\kom}$ introduced above \cite{zcontrolled}. 

In conclusion, we have shown
that the effective low-energy action 
(or Hamiltonian) of a ladder {with weak interaction} 
 is asymptotically \sof symmetric\cite{scalingmeans}.
With the inclusion of a
  next-nearest-neighbor hopping  $t_2$
the action is invariant under a generalized \soft transformation
 \cite{soft}, which
performs a ``stretched'' \sof rotation of the order parameters.
 Physically, this \soft symmetry 
 may be present in the low-energy sector of a {\it larger} and more
 {\it generic} class of
 physical systems than the ordinary \sof.
Moreover, 
since this 
 stretched rotation
does not conserve the norm of the superspin (order-parameter) vector\cite{zhan.97},
a renormalized
 \soft theory 
can possibly admit
 asymmetries between the
antiferromagnetic and superconducting phases, like for example the
difference in $T_c$'s
\cite{zhan.97,henl.97}.

We are grateful to  S. C. Zhang, B. Brendel, 
 and M. G. Zacher for many useful discussions. 
E. A. was supported by the
EC-TMR
  program  ERBFMBICT950048 and W. H. by FORSUPRA and BMBF (05 605 WWA 6).

\def\nonformale#1{#1}
\def\formale#1{}
\def\spa{} \def\spb{}
\def\andword{and }
%
%


%

%

%
%
%
%
%
%
%
%
%
%
%
%
%


\begin{thebibliography}{10}
\bibitem{soft}{
Throughout the paper, we denote by \soft the renormalized \sof symmetry,
  whenever the corresponding transformation \rsoft is expressed in terms of the
  original fermion fields $C$. The transformation \rsoft is obtained by first
  performing the $\other T$ transformation (described in the text), then the
  \sof transformation on the new fields\cite{scalingmeans}, and then
  transforming back to the original fermion fields, i. e., $\rsoft = \other
  T^{-1} \rsof \other T$. To illustrate what invariance under \rsoft means,
  let's take the non-interacting case. The Greens function of the band $\ky=0$
  is $G(\iom,\kx,\ky=0)=1/(\iom - V_0 \kx)$. We choose a particular \sof
  transformation for \rsof, namely, the exchange of the two bands. We thus have
  $\rsoft G(\iom,\kx,0) = \other T^{-1} \rsof V_0^{-1/2} G(\iom/\sqrt{V_0},\kx
  \sqrt{V_0},0)= \other T^{-1} V_0^{-1/2} G(\iom/\sqrt{V_0},\kx
  \sqrt{V_0},\ky=\pi)= (V_0/V_{\pi})^{-1/2} G(\iom/\sqrt{V_0/V_{\pi}},\kx
  \sqrt{V_0/V_{\pi}},\pi)= 1/(\iom - V_0 \kx) $. Our RG calculation
  additionally shows that
  the invariance under \rsoft holds (asymptotically) also for the interacting
  case, which is a non-trivial result.}%
\bibitem{scalar}{
One can show that it is not necessary to introduce tensor interactions for the
  \sof-invariant action (cf. \cite{notation}), as they are already included in
  the scalar form \eqref{si} due to the Fierz identity (see also Ref.
  \cite{sc.zh.97}).}%
\bibitem{coupl}{
The couplings depend on $(\rr_1,\ky_1;\rr_2,\ky_2|\rr_3,\ky_3;\rr_4,\ky_4)$,
  and are defined in analogy with the g-ology of Ref.\cite{soly.79}, as
  $g_i\typ4 \equiv$ \mbox{$g_i(+,0;+,0|+,0;+,0)$}, $g_i\typ{4;0-\pi} \equiv$
  \mbox{$g_i(+,0;+,\pi|+,\pi;+,0)$}, for the intra- and inter-band
  parallel-spin, $g_i\typ{2} \equiv$ \mbox{$g_i(+,0;+,0|-,0;-,0)$},
  $g_i\typ{2;0-\pi} \equiv$ \mbox{$g_i(+,0;+,\pi|-,0;-,\pi)$}, for the forward,
  and $g_i\typ{1} \equiv$ \mbox{$g_i(+,0;-,0|-,0;+,0)$}, $g_i\typ{1t;0-\pi}
  \equiv$ \mbox{$g_i(+,0;-,\pi|-,0;+,\pi)$}, and $g_i\typ{1;0-\pi} \equiv$
  \mbox{$g_i(+,0;-,\pi|-,\pi;+,0)$} for the backward couplings, where $i=0$,
  $1$, or $ph$ (see also text and \cite{web}).}%
\bibitem{scaling}{
In order not to confuse the reader, we have chosen for this new transformation
  the same symbol ``\ $\wtilde{ }$\ '' as for the one performed in the
  procedure (i), since they have the same physical meaning. The two
  transformations differ for large $\flow$ by (a) an overall constant, and (b)
  by a scaling of the frequencies and of the momenta. Notice, however, that due
  to their frequency- and momentum-independence, the $\wtilde g_i\tgen$
  couplings defined in (ii) are equivalent to the $\other g_i\tgen$ defined in
  (i) apart for a common factor.}%
\bibitem{approx}{
We have verified that, as a function of $t_2$ and for different $U\ $, $\other
  g_1\typ{max}/\other g_0\typ{max} \to \approx 0.1\ t_2^2$.}%
\bibitem{remaining}{
The only remaining explicitly non-\sof invariant term in the renormalized
  action is due to the different momentum and frequency conservation for the
  $\ky=0$ and $\ky=\pi$ bands arising from the $\other T$ transformation.
  However, 
one can show that this term
  also breaks the \sof symmetry by an amount of order $t_2^2$ at most.}%
\bibitem{zcontrolled}{
{ In the case of the ladders, where $Z_{\ky\flow}$ vanishes at $\flow_c$ due to
  a spin gap, this argument can be applied up to, say $\flow=\flow^*$ 
for which
  $Z_{\ky\flow}$ is still appreciable and still the \sof symmetry is achieved
  (see \onlinecite{controlled}) }.}%
\end{thebibliography}
\end{document}